\documentclass[12pt,preprint,a4paper]{aastex}

\usepackage{graphics,epsf}
\usepackage{amsmath}                
\usepackage{amsfonts}               
\usepackage{amssymb}                
\usepackage{epsfig}                 

\def \cm{~\rm{cm}}
\def \s{~\rm{s}}
\def \km{~\rm{km}}

\def \K{~\rm{K}}
\def \g{~\rm{g}}

\def \AU{~\rm{AU}}

\def \yr{~\rm{yr}}

\begin{document}

\title{A tidally destructed massive planet as the progenitor of the two light planets around the sdB star KIC 05807616}

\author{Ealeal Bear$^1$ and Noam Soker$^1$}

\affil{1. Department of Physics, Technion -- Israel Institute of
Technology, Haifa 32000, Israel;
ealeal@physics.technion.ac.il; soker@physics.technion.ac.il}

\begin{abstract}
We propose that the two newly detected Earth-size planets around
the hot B subdwarf star KIC 05807616 are remnant of the tidally
destructed metallic core of a massive planet. A single massive
gas-giant planet was spiralling-in inside the envelope of the red
giant branch star progenitor of the extreme horizontal branch
(EHB) star KIC 05807616. The released gravitational energy unbound
most of the stellar envelope, turning it into an EHB star. The
massive planet reached the tidal destruction radius of $\sim 1
R_\odot$ from the core, where the planet's gaseous envelope was
tidally removed. In our scenario the metallic core of the massive
planet was tidally destructed into several Earth-like bodies
immediately after the gaseous envelope of the planet was removed.
Two, and possibly more, Earth-size fragments survived at orbital
separations of $\ga 1R_\odot$ within the gaseous disk. The bodies
interact with the disk and among themselves and migrated to reach
orbits close to a 3:2 resonance. These observed planets can have a
planetary magnetic field about ten times as strong as that of
Earth. This strong magnetic field can substantially reduce the
evaporation rate from the planets and explain their survivability
against the strong UV radiation of the EHB star.
\end{abstract}


\section{Introduction}
\label{sec:intro}
Two planets around the sdB star KIC 05807616 (also known as KPD 1943+4058) were
recently announced by Charpinet et al. (2011).
The star is an extreme horizontal branch (EHB) star, that burns helium in its core and contains a very small
amount of mass in its envelope.
Spectroscopically, EHB stars are classified as hot subdwarf (sdB or sdO) stars, although some sdB and sdO stars might be post
asymptotic giant branch (AGB) stars.
In what follows we will refer by EHB stars to both groups (sdB and sdO).
The planets` deduced orbital separations are $a_{\rm P1}=1.29R_\odot$ and $a_{\rm P2}=1.64R_\odot$, for KOI 55.01 and
KOI 55.02, respectively. Their orbital periods are close to a 3:2 resonance (more accurately, 10:7 resonance).
Their respective masses are $0.44M_\oplus$ and $0.655M_\oplus$.
These small orbital separations imply that the progenitor of the KIC 05807616 planetary system
went through a common envelope (CE) phase, where a lower mass companion
spiraled inside the bloated envelope of the red giant branch (RGB) stellar progenitor and ejected its envelope.

In order to become an EHB star, the RGB progenitor must lose most
of its envelope (Dorman et al. 1995; Fusi Pecci et al. 1996). The
cause of this massive mass-loss is an unsolved issue in stellar
evolution. The debate is whether a single star (e.g., Yi 2008) can
account for the formation of hot subdwarfs, or whether a binary
evolution is behind the hot subdwarf phenomenon (e.g., Han et al.
2002, 2007). The finding that about half of the sdB stars in the
field (not in globular clusters) reside in close binaries with
periods as short as one day or less (Maxted et al. 2001;
Napiwotzki et al. 2004; Han \& Podsiadlowski 2008;  Copperwheat et
al. 2011) supports the binary model. Most of the formation
channels of sdB stars are summarized by Han et al. (2003),
although they omit the substellar channel.

It seems that EHBs can also be formed by interaction with
substellar companions (Soker 1998). In this scenario planets
enhance the mass-loss rate on the RGB and lead to the formation of
EHB stars. Massive planets might even survive the CE phase. EHB
stars with substellar companions at separations of $a_f \ga 1 \AU$
have also been found (Silvotti et al. 2007; Lee et al. 2009; Qian
et al. 2009; 2012). In these systems the wide substellar
companions might hint on a closer planet that went through the CE
phase and ejected the envelope of the RGB stellar progenitor of
the EHB star. This closer planet might have been completely
destructed in the CE.

Charpinet et al. (2011) suggest that the two planets they have discovered are descendants of two
massive planets that were already in a 3:2 resonance before entering the CE envelope with the RGB
progenitor of the EHB star. According to that suggestion, the two planets maintained this resonance during the CE phase.
The envelopes of the massive planets were evaporated in the CE, leaving behind their two respective metallic cores.
We find it unlikely that two planets maintained their resonance when entering a CE phase.
Necessarily one planet is engulfed first, and since the dynamical friction time in the
envelope is shorter than the gravitational interaction time
between planets, the resonance will be lost.
In addition, when the second planet enters the envelope, the envelope mass between them will be several times their combined masses.
The envelope will `screen' any mutual influence between the planets. The resonance cannot hold itself.
Even before entering the CE we expect the resonance to be lost due to the strong tidal interaction of the
closer planet with the RGB envelope.
Over all, we expect the closer planet to reach a distance of $\sim 1 R_\odot$ much
before the second planet. Either the envelope is ejected and the second planet ends at a larger distance than observed,
or the closer one collides with the core.
But we don't expect both planets to be so close to the core and to each other.

We here suggest an alternative scenario where these two planets
are the remnant of the tidally destructed core of a single massive
planet. The envelope of the destructed planet formed a gaseous
disk. The small planets migrated somewhat inside this disk to
enter the resonance. The basic scenario is described in section 2.
The survivability of the present Earth-like planets to evaporation
is discussed in section 3. Our short summary is in section 4.

\section{Tidal destruction}
\label{sec:Tidal}

In our proposed scenario a single massive gas-giant planet spirals
inside the RGB stellar envelope down to the tidal radius $R_t$
where it is destructed. To reach that radius the massive gas-giant
planet must escape evaporation due to the high temperature of the
RGB envelope. The approximate orbital separation at evaporation is
taken from Soker (1998) $a_{\rm EVA} =(m_{\rm P}/10 M_{\rm
J})^{-1} R_\odot$, where $m_{\rm P}$ is the planet mass and
$M_{\rm J}$ is Jupiter mass. This limits the substellar object in
our proposed scenario to be of a mass of $m_{\rm P} \ga 5 M_{\rm
J}$.
{{{

We will use the tidal radius both for the massive planet and for
its metallic core. The tidal radius depends on the destructed
object, e.g., its spin rate and equation of state through a
coefficient $C_{\rm tide}$ (e.g., Harris 1996; Davidsson 1999)
\begin{equation}
R_t \simeq C_{\rm tide} R_{\ast} \left( \frac{\rho_{\ast}}{\rho_{\rm P}} \right)^{1/3}
\simeq 1.8   
\left(\frac{C_{\rm tide}}{2} \right)
\left(\frac{M_{\ast}}{0.5 M_\odot}\right)^{1/3} \left(\frac{\rho_{\rm P}} {1 \g \cm^{-3}} \right)^{-1/3} R_\odot,
\label{Eq.Rt}
\end{equation}
where $R_{\ast}$ and $M_{\ast}$ are the radius and mass of the EHB star, respectively,
and $\rho_{\rm P}$ is the density of the planet.
Harris (1996) summarized the values of $C_{\rm tide}$ according to the properties of the destructed object; they
are in the range $\sim 1.3- 2.9$.


With these uncertainties, and with a similar density of the entire
massive planet and its metallic core, $\rho_{\rm P} \simeq
\rho_{\rm core} \simeq 5 \g \cm^{-3}$, their tidal radius is very
similar at $R_t \simeq 1 R_\odot$. As soon as the gaseous part is
tidally destructed, so does the core. The massive planet is
expected to have an eccentric orbit at the last phases of the
orbital phase (Taam \& Ricker 2006; Ricker \& Taam 2012), and
hence the destruction of the core will come immediately after the
destruction of the envelope even if the core is somewhat denser
than the planet. While the destruction of the gaseous envelope
will form a gaseous accretion disk, the destruction of the core
might leave several smaller Earth-like planets. Tidal destruction
results in an energy distribution among the destructed segments
(Lodato et al. 2009). Therefore, some of the newly formed
Earth-like planets will spiral further in an be completely
destructed, while some others will move outside the radius $R_{t}$
and survive. Two of these fragments, we propose, are the planets
discovered around KIC 05807616 by Charpinet et al. (2011).
Charpinet et al. (2011) mentioned the slight possibility of a
third small body in the system. Our scenario can account for a
third body if presence in the system.

The two remnants will migrate inside the gaseous disk to establish
a resonance (or almost a resonance), as is often claimed for
planets around young stars (e.g., Plavchan \& Bilinski 2011; Hasegawa \&
Pudritz 2011 and references therein). A more definite acceptance
of this scenario requires 3D hydrodynamical simulations of the
tidal-destruction process.

\section{Surviving evaporation}
\label{sec:Evap}

The Earth-like planets are exposed to intense UV radiation from
the central EHB star. This radiation evaporates the outer layer of
the planets (this is a different mechanism than the evaporation of
the massive planet inside the RGB envelope mentioned in section
\ref{sec:Tidal}). To estimate the evaporation rate we follow the
calculation of the evaporation rate by ionization of Jupiter-like
planets around EHB stars, but adopt the parameter to Earth-like
planets. We note that Rappaport et al. (2012) assumed that the
evaporation of the putative Mercury-like planet in KIC 12557548 is
driven by a thermal wind. In KIC 12557548 the central star is a
cool main sequence star ($T_{\rm eff} \simeq 4400 \K$), while in
our case the star is hot with intense UV radiation. For that we
use an evaporation mechanism based on ionization.

The evaporation rate is given by (Bear \& Soker 2011b)
\begin{equation}
\dot m_{\rm P} \simeq  2\pi c_s \mu m_H R_{\rm p}^{1.5}
a_{\rm P}^{-1}\sqrt{\frac{\tau N_* \eta_i}{8 \pi }}.
\label{Eq.dmdt3}
\end{equation}
where $\tau/n$ is the recombination time, $n$ is the total number
density of the ablated layer, $R_{\rm P}$ is the planet's radius,
and $c_s \simeq 2 \km \s^{-1}$ is the sound speed taken at the
atmosphere temperature of $T =9000K$ (Charpinet et al. 2011) and
for a singly ionized iron $\mu m_H \simeq 28 m_H$ (using silicon
will not change the results much). The rate of ionizing photons
hitting the planet $N= N_* (R_{\rm P}/{2a_p})^2$ is calculated
from the temperature and luminosity of KIC 05807616 ($T_{\rm eff}=
27,730 \K$ and $L= 22.9 L_\odot$) that give $N_* = 5 \times
10^{44} \s^{-1}$ for the number of ionizing photons per unit time
emitted by the EHB star. Here $\eta_i \simeq 0.1$ is the
ionization efficiency.

Substituting the above listed values in equation (\ref{Eq.dmdt3}) we find
\begin{equation}
\dot m_{\rm P} \simeq  10^{13}
\left(\frac{R_p}{0.8R_\oplus}\right)^{1.5}
\left(\frac{a_p}{1.3R_\odot}\right)^{-1}\left(\frac{\tau}{10^{12}
\s \cm^{-3}}\right)^{\frac{1}{2}}\g \s^{-1},
\label{Eq.dmdt4}
\end{equation}
where $\tau$ is calculated according to data in Osterbrock (1989).
The evaporation time of the close planet is $\tau_{\rm eva} \simeq
10^7 (\dot m_{\rm p} / 10^{13} \g \s^{-1})^{-1} \yr$. This
evaporation time is about 10\% of the HB life time.

However, the practical evaporation time might be much longer.
First, the evaporation rate derived above is highly uncertain, and
we might overestimate it by a factor of a few. Second, it is quite
plausible that a planetary magnetic field suppresses the
evaporation (Barnes et al. 2010; Haghighipour 2011 and references
therein). For that to occur the magnetic pressure should be about
equal or larger than the ram pressure of the outflowing gas
$B_{\rm P}^2/8 \pi \ga \rho v^2$, where for the outflow velocity
near the surface we take the sound speed and the density is given
by $\rho=\dot m_p / (2 \pi R_p^2 v)$ for an outflow in the half
sphere facing the central star. For the evaporation derived above
the constraint on the planet magnetic field to suppress the
outflow is
\begin{equation}
B_{\rm P} \ga \left( \frac {4 \dot m_p  v}{R_{\rm P}^2} \right)^{1/2} \simeq 5.5 \left( \frac {\dot m_p}
{10^{13} \g \s^{-1}} \right)^{0.5} \left( \frac {R_p}
{0.8R_\oplus} \right)^{-1} \left( \frac {v} {2 \km \s^{-1}}
\right)^{1/2} {\rm G}
\label{eq:ramp1}
\end{equation}
This is about ten times the Earth magnetic field.
Considering that the planets have suffered a recent strong tidal deformation with continues
strong tidal force from the central EHB star, and that they are heated by the central star radiation,
they are expected to be hot and liquid, and posses a differential rotation.
It is very probable that a strong dynamo is operating in each planet, leading to the required magnetic field.
Our prefer explanation for a low evaporation rate is the existence of a magnetic field.

\section{Summary}
\label{sec:summary}

The existence of two close Earth like planets (Charpinet et al.
2011), with $(M_{\rm P}, a_{\rm P})=(0.44M_\oplus, 1.29R_\odot)$
and $(0.655M_\oplus, 1.64R_\odot)$, around an EHB (sdB) star
raises the questions of their formation and survivability. The
mere presence of planets around EHB stars is not new (Silvotti et
al. 2007; Lee et al. 2009; Qian et al. 2009, Geier et al. 2009 but
see Jacobs et al. 2011 who argue that no planet exist in HD
149382) and not surprising (Soker 1998). What is surprising is
that these are Earth-like planets and that both are very close to
the central star. In their discovery paper Charpinet et al. (2011)
suggested that these two planets were originally two more massive
planets that orbited the RGB stellar progenitor. They both enter
the common envelope (CE) phase with the RGB star, spiraled all the
way to the center, and evaporated, leaving behind there metallic
cores. Their close to 3:2 orbital resonance have been maintained
during the CE phase. In section 1 we discuss why we find this
process unlikely.

We suggest the following alternative scenario for the formation of
the two planets (section 2). A single massive planet of mass
$m_{\rm P} \ga 5 M_{\rm J}$ went through the CE evolution inside
the RGB envelope. It spiralled all the way to the center. The
released gravitational energy is behind the removal of the stellar
envelope, as is commonly the case with CE evolution. The massive
planet reached the tidal destruction radius (eq. \ref{Eq.Rt}). The
gaseous mass of the planet was lost and part of it formed a
temporary accretion disk around the core, that now is the EHB
star. The metallic core of the massive planet was tidally
destructed into several Earth-like bodies immediately after the
gaseous envelope was removed. Different bodies had different
energy per unit mass. Some of them spiral-in and were further
destructed by the core, while other survived at orbital
separations of $\ga 1R_\odot$ within the gaseous disk. The bodies
interacted with the disk and among themselves and migrated, as
planets around young stars do. Two of the bodies survived and
reached an almost resonance. These are the observed Earth-like
planets.

The future of these planets is a direct result of the
evolution of the EHB. If the radius of the EHB will increase it
might engulf the planets prior to the formation of the WD.
We assume that the metallicity of KIC 05807616 is similar to solar
metallicity. Following Dorman et al. (1993, see Fig. 3d there)
evolutionary track for $L_{\rm EHB}\sim 22.9L_\odot$ and $T_{\rm
eff}\sim 27730K$ (Charpinet et al. 2011), it appears that the
radius will not increase much (or even not at all), and is not supposed to exceed
$R\sim 0.3R_\odot$. Therefore the planets are likely to remain in
orbit around the descendant CO WD.

 In section 3 we examined the survivability of the planets to evaporation by the UV radiation of the EHB star.
Equation (\ref{Eq.dmdt4}) for the evaporation rate implies that
the inner planet will be completely evaporated within $\sim 10^7
\yr$. This is shorter than the $\sim 10^8 \yr$ life duration on the HB.
However, the expression is highly uncertain, and we might overestimate the evaporation rate.
More likely we find the possibility that a planetary magnetic field, about ten times as strong as that of Earth
(eq. \ref{eq:ramp1}), will substantially reduce the evaporation rate by
holding the ionized gas.

This research was supported by the Asher Fund for Space Research at the Technion, and the Israel Science Foundation.


\end{document}